\documentclass[11pt]{article}

\usepackage[utf8]{inputenc}
\usepackage{amssymb,verbatim,epsfig}
\usepackage{amsmath,setspace}
\usepackage[amssymb]{SIunits} % SI-units
\usepackage{multicol}
\usepackage{multirow}
\usepackage{lipsum}  
\usepackage[style = numeric-comp, sorting = none, backend=biber,maxbibnames=6,doi=true, eprint=false]{biblatex}
\DeclareSourcemap{
  \maps[datatype=bibtex]{
    \map{
      \step[fieldset=issn, null]
      \step[fieldset=arxiv, null]
    }
  }
}

\usepackage[usenames,dvipsnames]{xcolor} % color package
\definecolor{MyBlue}{rgb}{0,0,0.45} %color used for clickable links and so on
\usepackage{url} %Is added
\usepackage[breaklinks,colorlinks=true, urlcolor=MyBlue, linkcolor=MyBlue, plainpages=false, citecolor=MyBlue,bookmarks=true,bookmarksopen=true,bookmarksnumbered=true]{hyperref}
\usepackage{breakurl} 
\usepackage{graphicx}
\usepackage{longtable}
\usepackage{lineno}
\usepackage[margin=1in]{geometry} 
\usepackage[small,bf]{caption}

\newenvironment{supplement}{
		\setcounter{table}{0}
		\setcounter{section}{0}
\setcounter{figure}{0}
}

\spacing{1} 
\usepackage{appendix}

\title{\bf 
\vspace{-0.75in}
Laser beam properties and microfluidic confinement control thermocavitation} 

\author{Jelle J. Schoppink$^{1*}$,
Jose A. Alvarez-Chavez$^{2}$,
David Fernandez Rivas$^1$ \\ 
\small$^1$Mesoscale Chemical Systems group, MESA+ Institute and Faculty of Science and Technology, \\  
\small University  of Twente, P.O. Box 217, 7500 AE Enschede, the Netherlands. \\
\small$^2$Optical Sciences group, MESA+ Institute and Faculty of Science and Technology, \\  
\small University  of Twente, P.O. Box 217, 7500 AE Enschede, the Netherlands. \\
\small $^*$ Corresponding author: j.j.schoppink@utwente.nl}

\date{\today}

\begin{document} \maketitle

\begin{abstract}
Thermocavitation, the creation of a vapor bubble by heating a liquid with a continuous-wave laser, has been studied for a wide range of applications. Examples include the development of an actuator for needle-free jet injectors, as the pumping mechanism in microfluidic channels and crystallization or nanoparticle synthesis. Optimal use in these applications require control over the dynamics of the laser-generated bubble through the laser power and beam radius. In contrast to pulsed lasers, for continuous-wave lasers the influence of the laser beam radius on the bubble characteristics is not fully understood. Here, we present a novel way to control the size of the beam from an optical fiber by changing the distance from the glass-liquid interface. We show that the increase in beam size results in a longer nucleation time. Numerical simulations of the experiment show that the maximum temperature at the moment of nucleation is 237$\pm$5\textdegree C and independent of laser parameters. Due to delayed nucleation for larger beam sizes, more energy is absorbed by the liquid at the nucleation instant. Consequently, a larger beam size results in a faster growing bubble, producing the same effect as reducing the laser power. We conclude that the total bubble energy only depends on the amount of absorbed optical energy and it is independent of the beam radius and laser power for any amount of absorbed energy. This effect contrasts with pulsed lasers, where an increase in beam radius results in a reduction of bubble energy. Our results are
of relevance for the use of continuous-wave laser-actuated cavitation in needle-free jet injectors as
well as other applications of thermocavitation in microfluidic confinement.
\end{abstract} 

\vspace{0.1 in}
\noindent{\bf Keywords:} [Thermocavitation, continuous-wave laser, energy transfer, microfluidic confinement, vapor bubble]

\vspace{0.5cm}

\section{Introduction}\label{sec: intro}

The creation of a vapor bubble by heating the liquid with a continuous-wave (CW) laser was first reported in 1987 by Rastopov and Sukhodol'skii, who called it thermocavitation~\cite{Rastopov1987}. Since then, this thermocavitation process has been studied for numerous applications, including removal of pathological tissues~\cite{Chudnovskii2017}, nanoparticle synthesis~\cite{Afanador-Delgado2022}, laser-induced crystallization~\cite{Korede2023}, as pumping mechanism in a microchannel~\cite{Garcia-Morales2021}, creation of short laser pulses~\cite{Zaca-Moran2021}, generation of ultrasound acoustics~\cite{Guzman-Barraza2022,Korneev2011,ZacaMoran2020b} and trapping or manipulation of bubbles~\cite{Sarabia-Alonso2020,Sarabia-Alonso2021}. Over the last decade, thermocavitation also has been investigated for its potential to create microfluidic jets for (bio-)printing and/or needle-free jet injection~\cite{Sopena2017,UrRehman2023,Schoppink2022,VanderVen2023,Schoppink2023-POF, Padilla-Martinez2013, Gonzalez-sierra2023}. Although laser-actuated jet injection was initially studied using pulsed lasers~\cite{Han2012,Tagawa2013,Krizek2020a,Krizek2020,Moradiafrapoli2017}, recently we found that continuous-wave lasers generate similar bubble dynamics~\cite{Schoppink2023-ETFS}. For all of these applications of thermocavitation, understanding and control of the bubble formation is vital. 

However, a thorough understanding of the thermocavitation process is still lacking.
Due to the low laser power (P~$\sim$~1W) in thermocavitation, the bubble does not form instantaneously upon laser irradiation, but after a short incubation time ($t_{n}\sim$~ms). Therefore, the delivered optical energy E$_{\textnormal{d}}$ is not controlled directly, but depends on the bubble nucleation instant (E$_{\textnormal{d}}$ = P$\times t_n$)~\cite{Schoppink2023-ETFS}. Delaying this nucleation time $t_n$ therefore increases the amount of energy, resulting in a larger bubble~\cite{Padilla-Martinez2014,Wang2018}. The most reported method for delaying the nucleation is a reduction in laser power~\cite{Ramirez-San-Juan2010,Padilla-Martinez2014}. Similarly, a reduction in absorption coefficient also reduces the absorbed energy density, which results in delayed nucleation, a larger superheated volume and consequently a larger bubble~\cite{Zhang2022,Afanador-Delgado2020}. A third method to deliver more energy and create a larger bubble is to increase the beam size. The beam size can be increased by moving the liquid away from the focal point of the focusing lens~\cite{Padilla-Martinez2011, Padilla-Martinez2014,Zhang2022, Gonzalez-sierra2023}. However, the exact beam size depends on the optics and positioning accuracy, which are difficult to reproduce. To our knowledge the quantitative influence of the beam size on the thermocavitation process and its energy transfer has not been reported. 

For thermocavitation, the temperature at the moment of nucleation is still debated. Fluorescent measurements using Rhodamine-B resulted in a maximum temperature of 98\textdegree C~\cite{Banks2019}. However, in this study the calibration was only performed until 85\textdegree C, and the sensitivity of this dye as temperature sensor goes down rapidly above 80\textdegree C~\cite{Ross2001}, for which reason any extrapolation to higher temperatures should be carefully interpreted. Numerical simulations resulted in temperatures of 295-332\textdegree C~\cite{Padilla-Martinez2014}, which is around or even above the spinodal temperature of 305\textdegree C~\cite{Caupin2006}, and therefore unlikely as nucleation at an interface should happen below the spinodal temperature~\cite{Avedisian1985}. 

In this manuscript, we investigate the influence of the beam size on the thermocavitation process in microfluidic confinement, i.e., near a wall boundary from which the laser is focused. We compare our experimental results on the bubble nucleation with a numerical heat transfer simulation in COMSOL. This data provides a better understanding of the moment of nucleation and the energy transfer from the CW laser into the bubble. Our results are of relevance for the use of continuous-wave laser-actuated cavitation in needle-free jet injectors as well as other applications of thermocavitation in microfluidic confinement. 

\section{Experimental methods}\label{sec: methods}

\begin{figure}[b!]
Figures    \includegraphics[width=\linewidth]{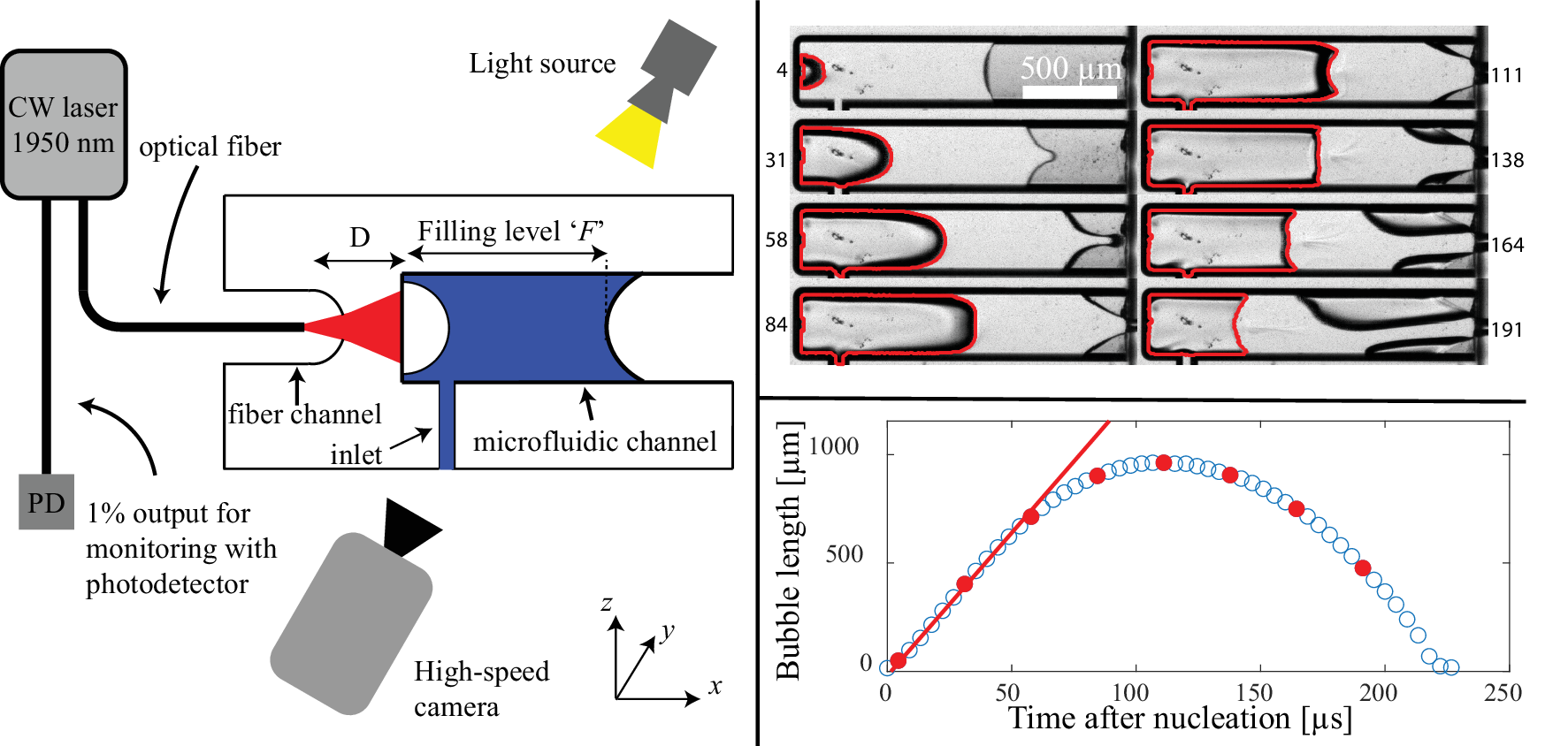}
    \centering
    \captionsetup{width=\linewidth}
    \caption{\textbf{Left:} Experimental setup consisting of a microfluidic glass chip with two etched channels. The microfluidic channel (right) is filled with water up  to distance $F$. The output fiber of a CW laser is inserted inside the fiber channel (left), and positioned with micrometer accuracy. The distance between the fiber tip and the microfluidic channel $D$ defines the beam size of the laser at the glass-liquid interface. Upon laser irradiation, nucleation occurs and a vapor bubble will grow and collapse. A high-speed camera with corresponding light source is positioned along the $y$-axis to capture $xz$-images. \textbf{Top right:} Eight selected images of the bubble in an experiment and its contour highlighted in red ($F$~=~1000~µm, P~=~550~mW, B$_{\textnormal{R}}$~=~18.7~µm). The numbers next to the frames indicate the time after nucleation in µs. Scale bar in first frame indicates 500~µm. \textbf{Bottom right:} Bubble length plotted versus time after nucleation. Lengths are calculated from bubble area inside the red contours divided by the channel height (400~µm). Red dots corresponds to eight frames above. Growth velocity is fitted from the first 5 frames, resulting in a slope of 13.2~m/s.}
    \label{fig: ExperimentalSetup}
\end{figure}

Figure~\ref{fig: ExperimentalSetup} shows the experimental setup, consisting of a microfluidic glass chip with two etched channels along the same axis and separated by 30~µm. The right channel (L*W*H~=~2000*100*400~µm$^3$) is partially filled with Milli-Q water until the variable filling level $F$. The left channel is designated for inserting an optical fiber connected to a CW laser. This single-mode optical fiber (Corning SMF-28e) is positioned inside its channel using a motorized 3-axis stage (Thorlabs Rollerblock) with micrometer accuracy. This allows for accurate aligning of the fiber tip with respect to the microfluidic channel. Due to the divergence of the laser beam from the fiber tip, the beam radius at the interface of the microfluidic channel can be controlled by the distance $D$. Seven beam radii B$_{\textnormal{R}}$ are used in the experiment between 10 to 36~µm.

The fiber laser (BKTel Photonics, HPFL-2-350-FCAPC) has a variable output power between 0.2 and 2~W at a wavelength of 1950~nm, which matches the absorption peak of water ($\alpha~\approx$~12000~m$^{-1}$~\cite{Ruru2012}). The laser has a secondary fiber output at 1\% of the nominal power, which is connected to a photodetector (Thorlabs DET05D2) to monitor the output power in-situ using an oscilloscope (Tektronix MSO2014B). 

Upon laser irradiation, the water inside the right channel is heated, and after a short period ($t_{n} \sim$~ms), nucleation occurs and a fast growing vapor bubble appears. A Photron NOVA SA-X2 high-speed camera was used in combination with a Navitar 12x zoom lens system and a Schott CV-LS light source for visualization of the bubble dynamics. The camera was used at a frame rate of 225k fps, a resolution of 384*96 and a pixel size of 5~µm. Figure~\ref{fig: ExperimentalSetup} (right panel) shows a few typical images during the bubble lifetime. The images were analyzed with a custom-made MATLAB algorithm, which tracks the bubble over time as shown in the red contours. The bubble length is calculated as the area enclosed in the red contour divided by the channel height (400~µm). The growth velocity is taken by fitting the bubble length of the second to the fifth frame.

The heating phase was simulated in COMSOL until bubble formation. First, using the ray optics module, the beam radius in the water channel is obtained for the different fiber positions used in the experiment. These beam radii are then used in the heat transfer module to simulate the heating. The energy absorption is calculated with Lambert-Beer, using the absorption coefficient of water, which reduces with increasing temperature~\cite{Jansen1994,Lange2002}. It also includes the loss of heat due to dissipation into the walls of the glass chip. More details regarding the numerical simulations of the ray tracing and heat transfer can be found in the Supplementary Information Sections~\ref{sec: suppl. ray tracing} and ~\ref{sec: suppl. heat transfer}, respectively.

\section{Results and discussion}\label{sec: results and discussion} 
\subsection{Nucleation time and temperature}\label{sec: Nucl time and temperature}

\begin{figure}[b!]
\begin{minipage}[t]{0.5\linewidth}
\centering
\includegraphics[width=\textwidth]{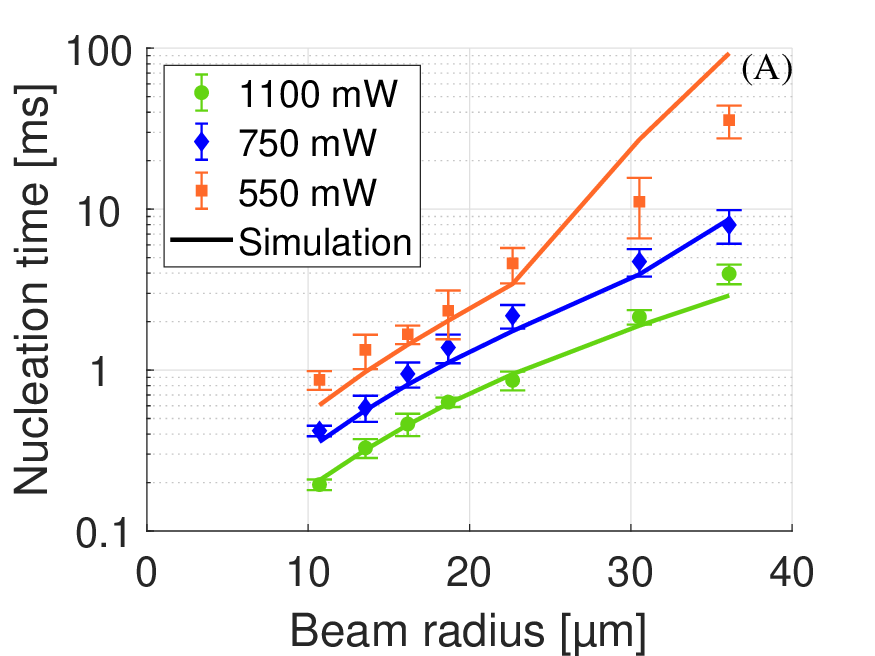}
\end{minipage}
\hspace{0.5cm}
\begin{minipage}[t]{0.5\linewidth}
\centering
\includegraphics[width=\textwidth]{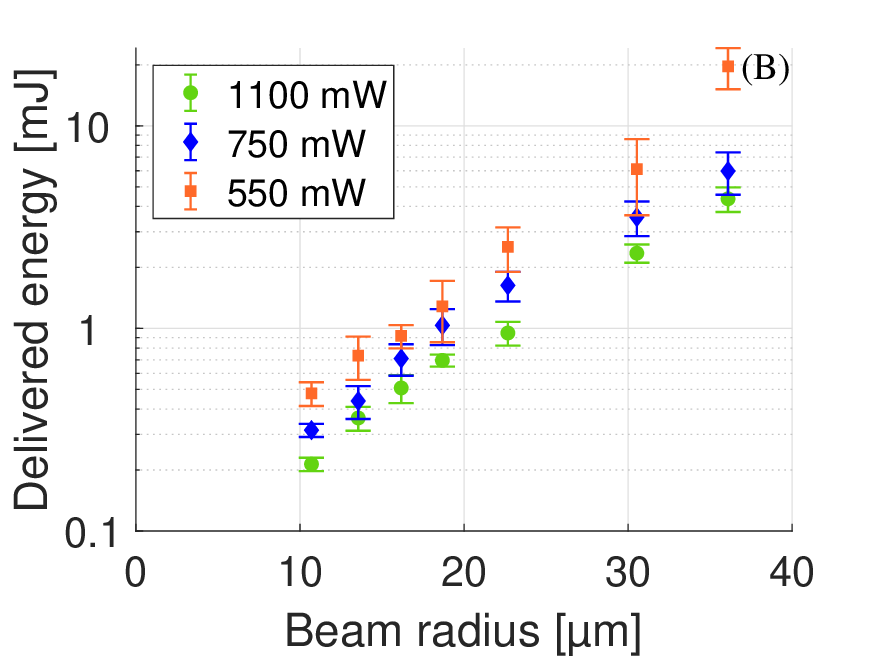}
\end{minipage}
\caption{Nucleation time (A) and delivered energy (B) as function of beam radius, for three different laser powers. In both figures, the error bars indicate the standard deviation for at least 6 individual experiments. The solid lines in (A) indicate the time in the COMSOL simulations at which the maximum temperature in the liquid is equal to 237\textdegree C. The delivered energy in (B) is calculated as E$_{\textnormal{d}}$~=~$\textnormal{P}_{\textnormal{L}}*t_{n}$.}
\label{fig: nucl_time_energy_beam_radius}
\end{figure}

Figure \ref{fig: nucl_time_energy_beam_radius}A shows the nucleation time as a function of beam radius, for 3 different laser powers. The data points are averaged over at least 6 individual measurements and the error bars indicate the standard deviation. It is clear that the nucleation time increases with increasing beam radius, as well as reducing laser power. These two effects reduce the laser intensity, resulting in slower localized heating of the liquid and therefore a longer nucleation time. 

For the middle laser power (750~mW), the experiment was performed for three different filling levels, $F$ = 600, 1000 and 1700~µm. It was found that the filling level did not have any significant effect on the nucleation time (see Figure~\ref{fig: sup: Nucleation times fill level}). This is explained as the filling levels are 8 to 20 times larger than the absorption length ($\delta \approx$~80~µm) and therefore the additional liquid has no effect on the heating, as all the optical power is absorbed before reaching the position of the meniscus for the smallest filling level ($F$~=~600~µm). The typical length over which heat diffusion takes place, $\delta$, is calculated as~\cite{Paltauf2003}
\begin{equation}\label{Eq: heat diffusion}
    \delta = \sqrt{4\kappa{t}_{n}},
\end{equation}
where $\kappa$ is the thermal diffusivity (0.14~mm$^{2}$/s for water). Even for the longest nucleation time (35~ms), $\delta=70$~µm, and therefore at least one order of magnitude smaller than the filling level. Therefore, it can be assumed that changes in the filling level do not affect the nucleation times.

In the numerical simulations using COMSOL, the energy absorption and heat transfer in the liquid experiment is simulated until the moment of nucleation, which we took from the experimental nucleation times. These experimental nucleation times are in agreement with simulated times with a maximum local temperature of 237\textdegree C, see solid lines in Figure~\ref{fig: nucl_time_energy_beam_radius}A. This agreement indicates that nucleation temperature is independent of laser beam radius and power.
However, the two lowest laser intensities show an exception; where the beam radius is large and laser power small. In such cases, the experiment gives a smaller temperature due to a larger heated region, which is up to 30 times larger compared to the other cases.
In such situations, nucleation may happen at lower temperatures, and thus shorter nucleation times. For all other data points, the temperature is very close to the 237\textdegree C, with a standard deviation of 5\textdegree C (see Figure~\ref{fig: sup: Nucleation temperatures}).
These temperatures are all well above the boiling temperature of water at atmospheric pressure (100\textdegree C), which is explained by the existence of an energy barrier for nucleation. Due to this energy barrier, higher temperatures are needed for bubble formation in microfluidic volumes on short timescales (ms). On the other hand, these temperatures are well below the spinodal temperature at atmospheric pressure (306\textdegree C~\cite{Caupin2006}), which is explained as the bubble forms at a wall, where the energy barrier for bubble formation is lower~\cite{Atchley1989}. Due to this energy barrier, the nucleation itself is a stochastic event~\cite{Padilla-Martinez2014,Caupin2006}, which could further explains the slight variations in nucleation time and temperature. Furthermore, impurities such as gas molecules also reduce the energy barrier and therefore nucleation temperature. 

\begin{figure}[t!]
    \includegraphics[width=\linewidth]{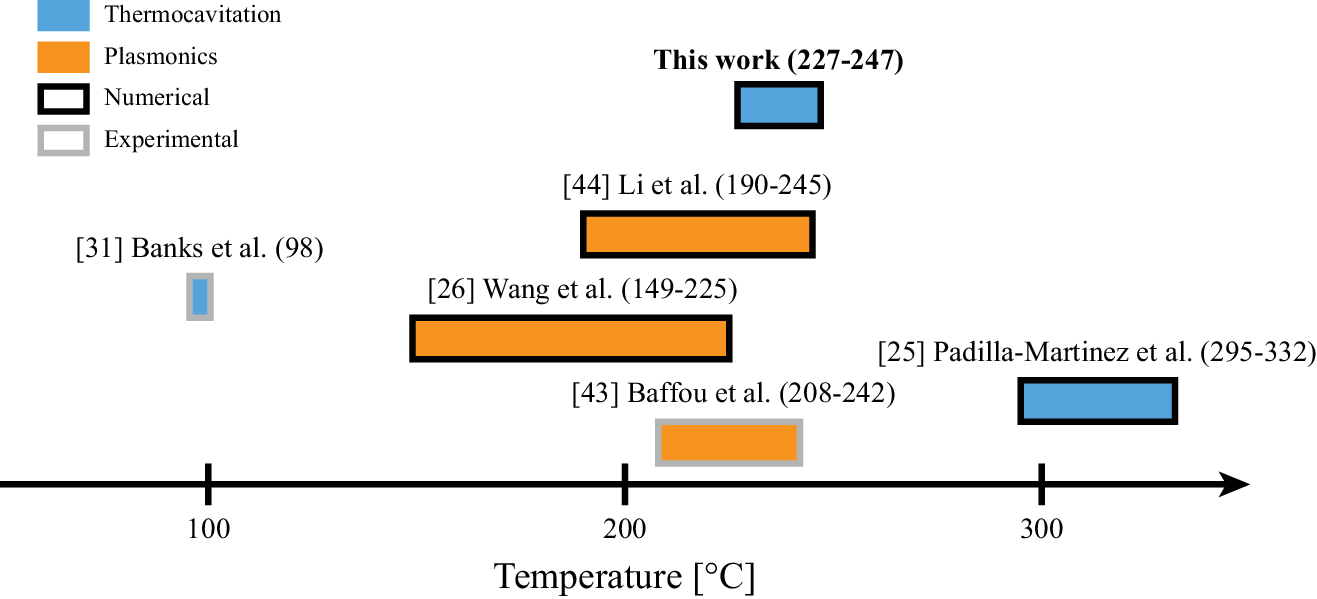}
    \centering
    \captionsetup{width=\linewidth}
    \caption{Calculated temperature ranges for nucleation in water with continuous-wave lasers found in literature. Blue colors indicate thermocavitation and orange plasmonic heating. Black outlines indicate numerical calculations, grey outline indicate experimental measurements. They appear in chronological order, with earliest work at the bottom.\nocite{Baffou2014}\nocite{Padilla-Martinez2014}\nocite{Wang2018}\nocite{Banks2019}\nocite{Li2019}
    %\cite{Baffou2014}\cite{Padilla-Martinez2014}\cite{Wang2018}\cite{Banks2019}\cite{Li2019}
    }
    \label{fig: nucleation temperatures}
\end{figure}

In literature, different temperatures are noted for bubble formation using a CW laser, either through thermocavitation (direct heating of the liquid) or plasmonic heating (indirect heating of the liquid through plasmonic nanoparticles), see Figure~\ref{fig: nucleation temperatures}. For thermocavitation, studies report different values, either close to the boiling temperature or the spinodal temperature, both of which are unlikely due to the above-mentioned reasons. Our values are in agreement to temperatures found for plasmonic bubbles, therefore, we conclude that our values are closer to the actual temperatures in thermocavitation. Nonetheless, heterogeneous nucleation depends on impurities, which act as nucleation sites~\cite{Caupin2006,Prosperetti2017}. These impurities, such as surface roughness~\cite{Atchley1989}, surfactants or dissolved gas molecules~\cite{Pettersen1994,Pfeiffer2022}, reduce the energy barrier and therefore result in earlier nucleation. 

Figure~\ref{fig: nucl_time_energy_beam_radius}B shows the delivered energy at the moment of nucleation as a function of beam radius, which is calculated by the laser power multiplied by the nucleation time (E$_{\textnormal{d}}$~=~$\textnormal{P}_{\textnormal{L}}*t_{n}$). We observe an increase of beam radius or decrease of laser power results in an increase in energy. Furthermore, E$_{\textnormal{d}}$ spans over two orders of magnitude (0.2~-~20~mJ), for a single thermocavitation set-up. Especially the beam radius plays a significant role in the delivered energy, which makes this set-up an optimal way to accurately control the amount of delivered energy.

\subsection{Bubble growth and energy conversion}

\begin{figure}[t!]
\begin{minipage}[t]{0.45\linewidth}
\centering
\includegraphics[width=\textwidth]{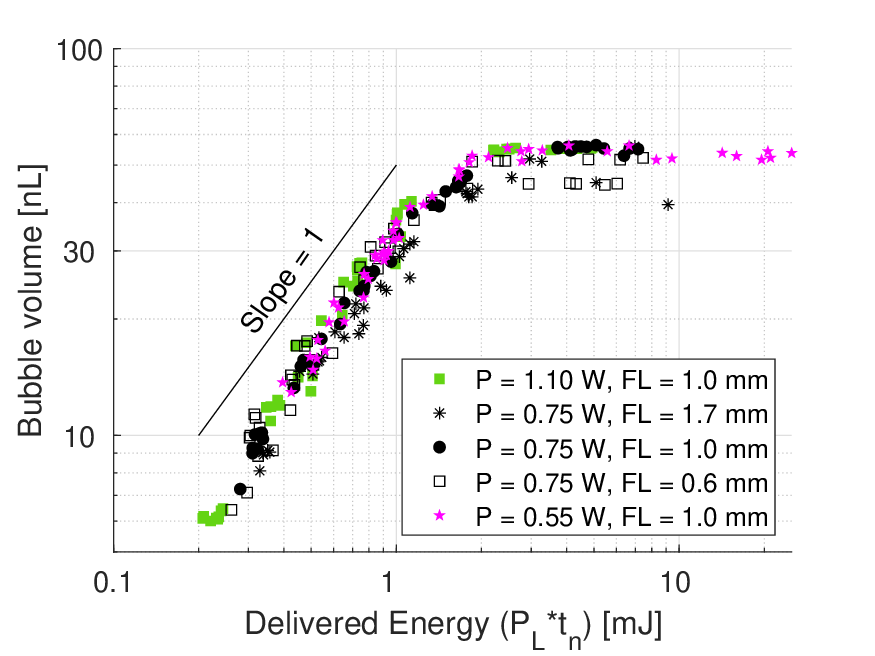}
\caption{Bubble volume vs delivered energy, for different laser powers P and filling levels $F$. Volume is calculated from bubble area multiplied by the channel depth (100~µm)}
\label{fig: Bubble Volume}
\end{minipage}
\hspace{0.5cm}
\begin{minipage}[t]{0.45\linewidth}
\centering
\includegraphics[width=\textwidth]{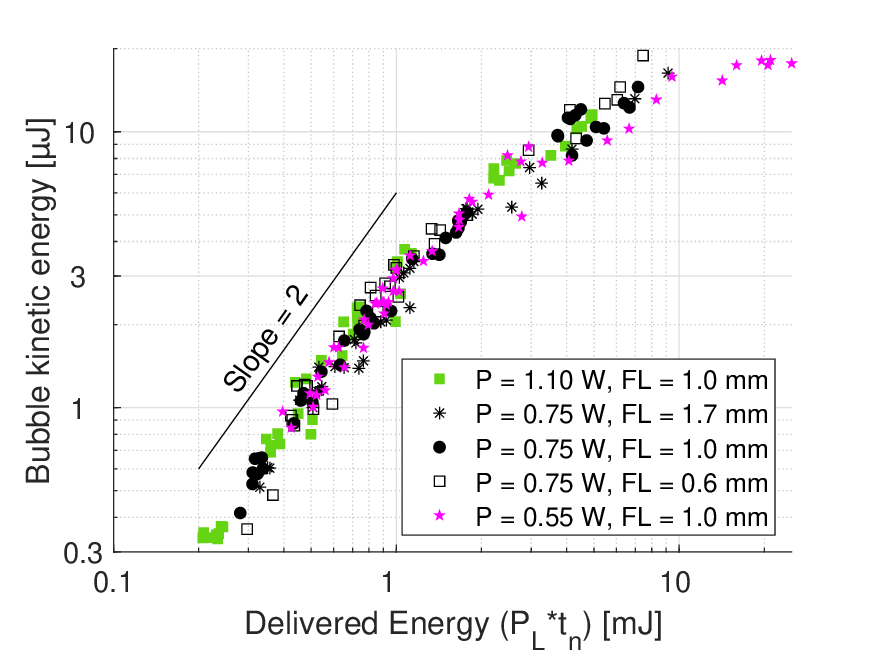}
\caption{Bubble kinetic energy vs delivered energy, for different laser powers P and filling levels $F$. Kinetic energy E$_{\textrm{kin}} = \frac{1}{2}$mv$^2$, where m is the liquid mass in the channel and v the maximum bubble growth velocity (change of length over per unit time)}
\label{fig: Kinetic energy}
\end{minipage}
\end{figure}

The maximum bubble volumes are shown in Figure~\ref{fig: Bubble Volume} as a function of delivered energy. For all experimental parameters, the maximum bubble volume increases linearly with delivered energy (see logarithmic slope of 1). As all data points are along the same curve, there is little influence of laser power or filling level. However, for large values of delivered energies (E$_{\textnormal{d}}\gtrsim$~2~mJ), the bubble volume plateaus. This plateau is explained by the limited channel length, as the bubble collapses at the moment they coalesce with the surrounding air (see example in Figure~\ref{fig: sup: Coalesce Collapse}). Therefore, the bubble never reaches its potential maximum volume and larger bubbles cannot be observed in this configuration. This is most apparent for the smallest filling level ($F=0.6$~mm, green diamonds), where already for smaller bubble volumes it can grow beyond the contact line and coalesce with the air inside the channel, resulting in lower plateau values in Figure~\ref{fig: Bubble Volume}.

Figure~\ref{fig: Kinetic energy} shows the kinetic energy of the bubble as a function of delivered energy. The kinetic energy is E$_{\textrm{kin}} = \frac{1}{2}$mv$^2$, where m is the liquid mass in the channel and v the maximum bubble growth velocity (change of length over per unit time). We note that the bubble kinetic energy increases quadratically (log slope = 2) with the delivered energy, independent of the laser or liquid parameters. This means that for a constant filling level, the bubble growth rate increases linearly with the delivered energy, as was also found in our earlier work~\cite{Schoppink2023-ETFS}. Here, we now also find that the mass (m~$\propto F$) does not affect the energy transfer and therefore the bubble growth rate v scales with v~$\propto F^{-0.5}$, which matches previous qualitative observations~\cite{OyarteGalvez2020}. For applications such as jet formation for printing or needle-free injection, this means that the liquid velocity can be controlled through the mass of liquid in the confinement. Furthermore, this independence of laser parameters contrasts with pulsed lasers, where an increase in beam radius results in a slower growing bubble~\cite{Krizek2020, Schoppink2023-ETFS}. For large values of the delivered energy, the slope in Figure~\ref{fig: Kinetic energy} decreases. This is explained by heat diffusion, as this large amount of energy is achieved through long nucleation times, at which point heat dissipation into the glass plays a significant role (see Equation~\ref{Eq: heat diffusion}). This is especially the case for the smallest laser power (P~=~0.55~mW, pink stars), which requires the longest nucleation times to reach those energies, resulting in more heat dissipation.

\begin{figure}[t!]
    \includegraphics[width=0.5\linewidth]{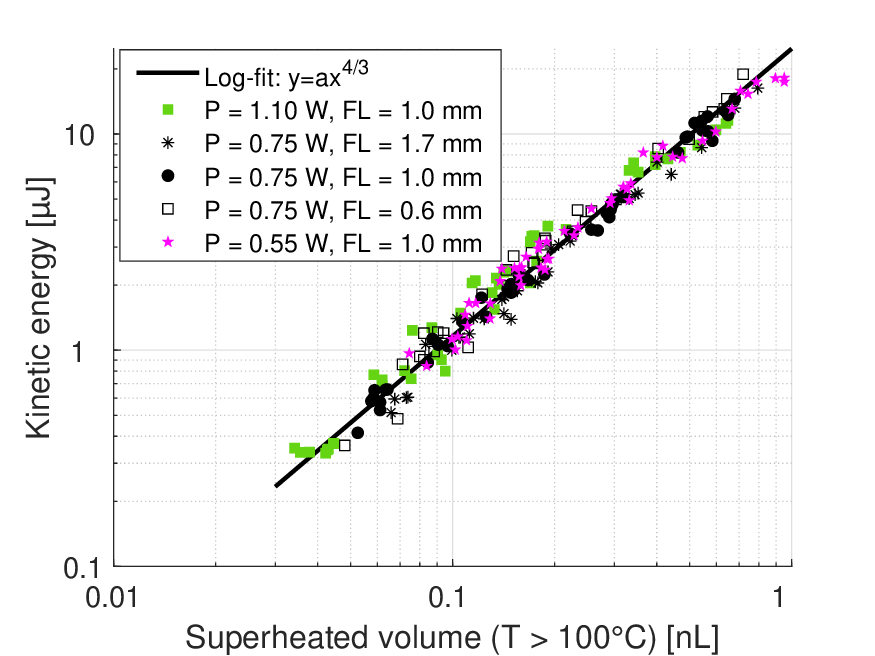}
    \centering
    \captionsetup{width=0.5\linewidth}
    \caption{Bubble kinetic energy vs volume of superheated liquid (T~$>$~100\textdegree C) at the moment of nucleation, taken from the simulation. Each data point is an individual experiment, where the color and symbol indicate the laser power and filling level. Solid black line is logarithmic fit with slope of 4/3.}
    \label{fig: Kinetic Energy vs Simulation Boiling volume}
\end{figure} 

One of the goals of the COMSOL simulations was to investigate the heat dissipation during the absorption of optical energy until moment of nucleation. As discussed in section~\ref{sec: Nucl time and temperature}, nucleation happens at approximately 237\textdegree C. However, after nucleation has occurred and the energy barrier has been overcome, liquid at a lower temperature (but still above 100\textdegree C) may also contribute to this growing bubble. Figure~\ref{fig: Kinetic Energy vs Simulation Boiling volume} shows the bubble kinetic energy as a function of the volume of superheated water (T~$>$~100\textdegree C) at the moment of nucleation. This superheated volume is taken from the COMSOL simulations at the moment of nucleation in the experiment. In contrast to Figure~\ref{fig: Kinetic energy}, where the initial quadratic relation seems to decrease, the slope in Figure~\ref{fig: Kinetic Energy vs Simulation Boiling volume} remains constant, which is explained as heat dissipation is included in this simulation. 

Microbubbles can also be created by different means, such as pulsed lasers~\cite{Schoppink2023-ETFS}, plasmonic bubbles~\cite{Li2019}, discharge with low-~\cite{Podbevsek2021} or high-voltage~\cite{GonzalezAvila2015}, microheaters~\cite{Nguyen2019} and the tube arrest method~\cite{Bao2023}. These methods can create similar bubble sizes as in this study and require similar amounts of energy~\cite{Schoppink2023-ETFS,Podbevsek2021}. Follow-up studies could focus on a quantitative comparison between the bubble dynamics and their R(t)-curves to find the best method for different applications. However, most of these methods are invasive, which reduces the ease of use and making chip fabrication more complex. The laser-generated bubbles allow for local heating and generation of bubbles on-chip, and more specifically the use of CW lasers allow for small and affordable set-up.

\section{Conclusion}\label{sec: conclusion}
We proposed and developed a novel set-up to accurately control the laser beam size for thermocavitation in microfluidic confinement. We compared experimental results using high-speed imaging to numerical simulations on the energy absorption and heat transfer. This study focused on the influence of laser beam characteristics on thermocavitation in microfluidic confinement and the energy conversion. We found that the nucleation time increases with increasing beam radius as well as decreasing laser power. Numerical simulations of the heat transfer show that the maximum temperature at the moment of nucleation is 237~$\pm$~10\textdegree C and independent of laser beam parameters. This temperature is below the spinodal temperature (306\textdegree C), but well above the boiling temperature (100\textdegree C) and is in agreement to earlier work on plasmonic bubbles. As the filling level $F$ is much larger than the absorption length, it does not influence the nucleation time or temperature.

Furthermore, we found that the maximum bubble volume increases linearly with delivered energy and the conversion is independent of laser parameters. For the largest energies, the maximum bubble volume reaches a plateau as the bubble coalesces with the surrounding air at the opening of the microfluidic channel before reaches its maximum potential volume. The bubble kinetic energy increases quadratically with the delivered energy. However, for large energies, the conversion efficiency decreases, which is explained by the heat dissipation, as the nucleation time is on the same timescale as thermal diffusion. From the temperature profiles in the numerical simulations we find that the bubble kinetic energy increases with volume of superheated liquid (T~$>$~100\textdegree C), with a power law of 4/3. As heat dissipation is included in these simulations, this relation holds for all data points, independent of the laser or liquid parameters.

Our findings contribute to the understanding and use of thermocavitation, and allow for a better control over the bubble characteristics in real life applications. The laser power and beam radius control the nucleation time and delivered energy, and can therefore control the bubble size and growth rate. This allows for optimal use of thermocavitation in a wide range of applications, including laser-actuated jet injection.

\section*{Acknowledgements}
J.J.S and D.F.R. acknowledge the funding from the European Research Council (ERC) under the European Union’s Horizon 2020 Research and Innovation Programme (Grant Agreement No. 851630). J.J.S. would like to thank Stefan Schlautmann for the fabrication of the microfluidic chips.

\section*{Competing interest}
The authors declare that they have no known competing financial interests or personal relationships that could have appeared to influence the work reported in this paper.

\section*{CRediT authorship contribution statement}
\textbf{Jelle J. Schoppink:} Conceptualization, Methodology, Formal analysis, Investigation, Data Curation, Writing - Original Draft, Visualization
\textbf{Jose A. Alvarez-Chavez:} Conceptualization, Writing - Review \& Editing.
\textbf{David Fernandez Rivas:} Conceptualization, Supervision, Project administration, Funding acquisition, Writing - Review \& Editing.

\printbibliography 

\begin{supplement}
%	\addtocontents{toc}{\protect\setcounter{tocdepth}{0}}
\newpage

\section{Numerical simulations: Ray tracing}\label{sec: suppl. ray tracing}
\subsection*{Geometry \& Material}

To calculate the size of the laser beam at the glass-water interface, ray tracing simulations were performed in COMSOL (version 6.0). 

The 3D geometry in the simulation is an exact copy of the microfluidic chip. Due to axial symmetry at $y=0$ and $z=0$, only a quarter of the domain has been simulated. The material properties of the air and the Schott BOROFLOAT \textregistered~33 glass were taken from the in-built materials. The water was taken as the in-built material from Hale and Quarry~\cite{Hale1973}. 

\subsection*{Calculation of initial beam divergence}
The initial beam divergence from the fiber tip in air can be calculated using fiber parameters. First, the V-value at our wavelength, which is the normalized frequency, is calculated~\cite{Senior2009}

\begin{equation}\label{Eq: V-value}
    V = \frac{2\pi}{\lambda}a\textrm{n}_{\textrm{core}}(2\Delta)^{\frac{1}{2}},
\end{equation}

where $\lambda$ is the wavelength (1.95~µm), a the fiber core radius (4.1~µm). $\Delta$ is the relative refractive index difference, calculated as
\begin{equation}\label{Eq: Delta}
    \Delta = \frac{\textrm{n}_{\textrm{core}}-\textrm{n}_{\textrm{clad}}}{\textrm{n}_{\textrm{core}}},
\end{equation}
where $\textrm{n}_{\textrm{core}}=1.45213$ and $\textrm{n}_{\textrm{clad}}=1.44692$ are the refractive index of the SMF28 fiber core and cladding respectively~\cite{Saktioto2020}, which results in a value of $\Delta = 0.0036$.

Using this value for $\Delta$ in Equation~\ref{Eq: V-value}, results in a V-value of 1.62. With this value, the mode field diameter (MFD) can be calculated~\cite{Senior2009}

\begin{equation}\label{Eq: MFD}
    \textrm{MFD} = 2a(0.65+1.619V^{-\frac{3}{2}}+2.879V^{-6}) = 12.8~\textrm{µm}.
\end{equation}

From the MFD, the Rayleigh range $z_{r}$ can be calculated by~\cite{SMF28}
\begin{equation}\label{Eq: Rayleigh}
    z_{r} = \frac{\pi}{\lambda}(\frac{\textrm{MFD}}{2})^{2} = 66~\textrm{µm}.
\end{equation}

From here, the beam radius $w(z)$ can be calculated as~\cite{SMF28}
\begin{equation}\label{Eq: BeamRadiusAtZ}
    w(z) = \frac{MFD}{2}\sqrt{1+(\frac{z}{z_{r}})^{2}}.
\end{equation}
In our experiments, $z$ equals to the the distance $D$, which is between 147-567~µm. This means that we are in the far-field as $(\frac{z}{z_r})^2>>1$. This allows combining and simplifying Equations~\ref{Eq: Rayleigh} and \ref{Eq: BeamRadiusAtZ} to
\begin{equation}
    w(z) = \frac{2\lambda}{\pi\textrm{MFD}}z.
\end{equation}
From here, the divergence can be calculated as 
\begin{equation}\label{Eq: BeamDivergence}
    \theta = \arctan{\frac{w}{z}} = \arctan{\frac{2\lambda}{\pi\textrm{MFD}}} = 0.097~\textrm{rad}.
\end{equation}

\subsection*{Ray modeling}
The laser beam was simulated as a Gaussian beam consisting of a total of 7651 individual rays, over a total of 50 polar angles. The initial divergence in air is 0.097~rad, as calculated in Equation~\ref{Eq: BeamDivergence}. They were released at the various $x$-positions of the fiber tip in the experiment. At the air-glass interface, each ray is refracted according to their incident angle. At the liquid-water interface, the rays were captured and their positions ($y,z$) and powers were saved to text files. Furthermore, 50 µm further into the channel, the rays were captured again, and their position and power were saved as well to quantify the effect of the divergence inside the water. In MATLAB, the individual rays were combined into a single curve for the incident power as a function of the radial position ($r = \sqrt(y^2+z^2)$). This was fitted with a Gaussian curve to get the standard deviation, which is close to the beam radius. The resulting beam radii can be found in Table~\ref{Table: fiber dists and sizes}.

To ensure that the time steps and mesh cells were sufficient small enough, the simulation was repeated for a range of time steps and meshes for one specific value of $D$ (440~µm) as can be seen in Table~\ref{table: ray refinement study}. It was found that time steps of $10^{-5}$~ns were sufficiently small enough, and the smallest mesh was chosen as it did not increase the computational time that much.

\begin{table}[h!]
\centering
\caption{Laser beam radius at the glass-water interface and halfway the absorption length in the water ($\approx$~50~µm) for the used distances $D$ in the experiment (distance between the tip of the optical fiber and the water channel). Beam radii are calculated with a ray tracing simulation in COMSOL.}
\label{Table: fiber dists and sizes}
\resizebox{\textwidth}{!}{%
\begin{tabular}{|l|l|l|l|l|l|l|l|}
\hline
Distance fiber tip to water channel (µm) & 147 & 193 & 236 & 276 & 339 & 470 & 567 \\ \hline
Laser beam radius at channel interface (µm) & 7.2 & 9.6 & 11.9 & 14.0 & 17.5 & 24.6 & 29.8\\ \hline
Laser beam radius 50 µm into the channel (µm) & 10.7 & 13.6 & 16.2 & 18.7 & 22.7 & 30.5 & 36.1\\ \hline
\end{tabular}%
}
\end{table}

\begin{table}[h!]
\centering
\caption{Beam radii for the mesh and time step refinement study. Columns are for constant time steps, rows for constant mesh element sizes, using the in-built physics-controlled mesh in COMSOL. The chosen settings are 'Extremely fine' and 1e-5, as indicated in red.}
\label{table: ray refinement study}
\begin{tabular}{|l|l|l|l|l|}
\hline
\begin{tabular}[c]{@{}l@{}}Time step (ns) /\\ Mesh element size\end{tabular} & 3e-6 & 1e-5 & 3e-5 & 1e-4 \\ \hline
Extremely fine &   24.67  & {\color{red}24.57} &  24.27  & 10.14\\ \hline
Extra fine &  24.67  & 24.57  & 24.27  &  9.55 \\ \hline
Finer &   24.67 &  24.56 &  24.27   & 9.51 \\ \hline
Fine &    24.74  & 24.64 &  24.34  &  9.55 \\ \hline
Normal &  24.68  &  24.58 &  24.28  & 9.12 \\ \hline
\end{tabular}
\end{table}

\section{Heat transfer simulations}\label{sec: suppl. heat transfer}
To simulate the heating of the liquid, the radiative beam in absorbing media module in COMSOL is used, in combination of the heat transfer in solids. The radiative beam is taken as a parallel Gaussian beam with the radii found in the earlier described ray tracing simulations (see Section \ref{sec: suppl. ray tracing}). As the beam is still expanding inside the water, this should be taken into account. To compensate for this effect, the laser beam radius is taken halfway across the absorption length, which is approximately 50~µm into the channel.

The absorption coefficient at the laser wavelength ($\lambda = 1.95$µm) is equal to $\alpha = 12121$ m$^{-1}$~\cite{Ruru2012}. However, it is found that the optical absorption of water in the 2~µm region decreases with increasing temperature~\cite{Vogel2003}, which is caused by the shift of the absorption peak towards shorter wavelengths~\cite{Jansen1994}. This decrease of absorption coefficient at the wavelength of 2.014~µm was found to be -25.9 K$^{-1}$m$^{-1}$~\cite{Lange2002}.
The optical absorption in the glass at the laser wavelength is negligible ~\cite{SCHOTT_optical} and therefore set to zero in the simulation. However, through heat dissipation from the water, the glass chip is heated, which results in a loss of energy conversion from the laser to the liquid.

As the channel is mirror symmetrical along $z=0$ and $y=0$, only a quarter of the channel is simulated. Due to the high absorption coefficient, the water channel is set to a length of 1000~µm, as no heat absorption or diffusion is expected to take place further into the channel. The time steps differ for each simulation, depending on the laser power and beam size, such that there are at least 300 time steps per simulation.

These simulations give the temperature profile at the moment of nucleation, which is taken from the experiment (see Figures~\ref{fig: nucl_time_energy_beam_radius}A~and~\ref{fig: sup: Nucleation times fill level}). From these temperature profiles (example shown in Figure~\ref{fig: sup: Temperature Profile}), we obtain the maximum temperature, which is always in the center of the laser beam, close to the wall, where the optical intensity is the highest. Furthermore, the temperature profile allows to calculate the volume of liquid which is superheated (T~$>$~100\textdegree C), which gives an indication of the liquid volume that can contribute to the bubble formation (see Figure~\ref{fig: Kinetic Energy vs Simulation Boiling volume}).

\begin{figure}[ht!]
    \includegraphics[width=\linewidth]{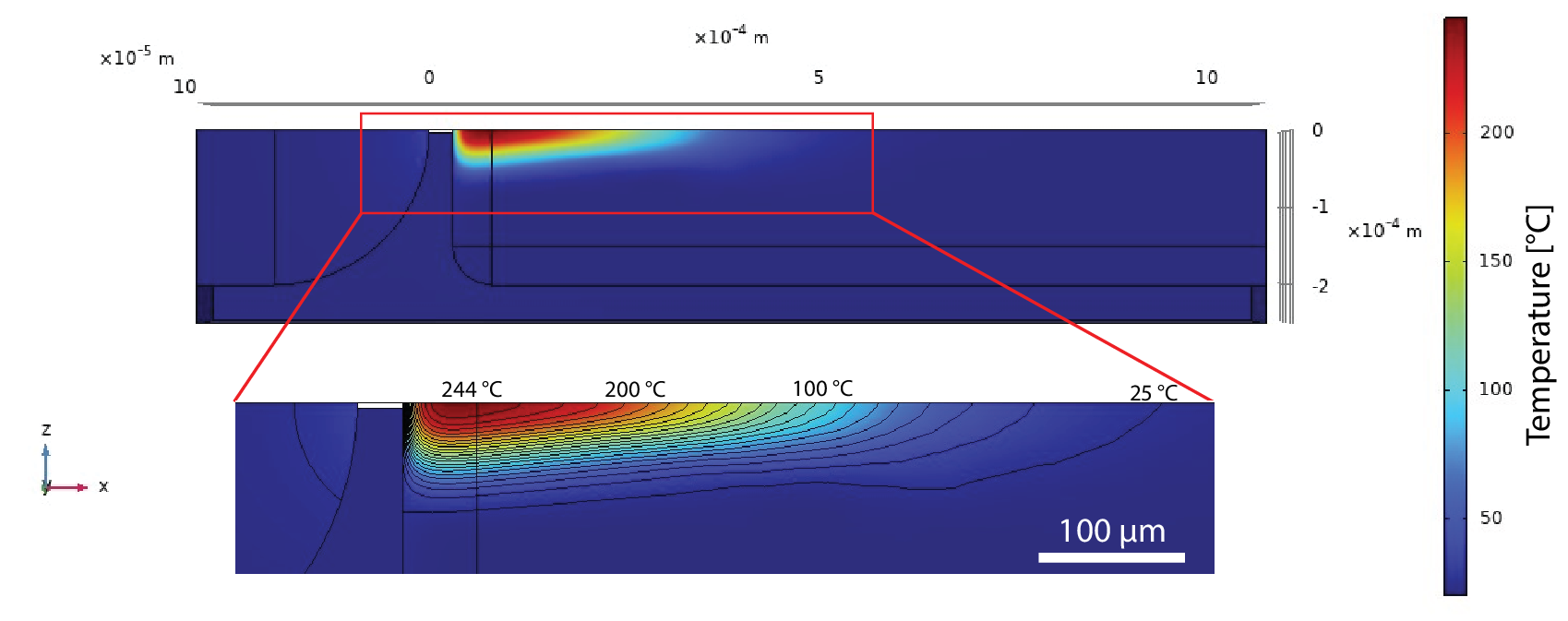}
    \centering
    \captionsetup{width=\linewidth}
    \caption{Example temperature profile of the $xz$-plane taken from the COMSOL simulation. Due to symmetry in the $z$-direction (at $z=0$), only half the channel is shown. The $xz$-plane is taken at the plane of symmetry, $y$~=~0. A close-up of the heated region is shown below, which spans approximately 500~µm into the x-direction, and 100 in the $z$-direction. The black lines indicate isotemperature lines, with an interval of 10\textdegree C, starting at 25\textdegree C. Beam radius is 18.7~µm, power is 750~mW, nucleation time (experimental) is 1.38~ms. From this profile, it is found that the maximum local temperature is 244\textdegree C, and the boiling volume is 1.98*10$^{-13}$~m$^{3}$, which is approximately equal to 200*30*30~µm$^{3}$.}
    \label{fig: sup: Temperature Profile}
\end{figure} 

\newpage
\section{Experimental nucleation times}\label{sec: suppl. exp nucleation times}
The experimental nucleation times for the different laser powers and filling levels are shown in Figure~\ref{fig: sup: Nucleation times fill level}. This Figure shows that the nucleation time only depends on the beam radius and power. For P~=~0.75W, three different filling levels $D$ are used. This filling level does not seem to influence the nucleation time, which can be expected as the filling level $D$ is much larger than the typical absorption length (1/$\alpha \approx$~80~µm). 
\begin{figure}[h!]
    \includegraphics[width=0.5\linewidth]{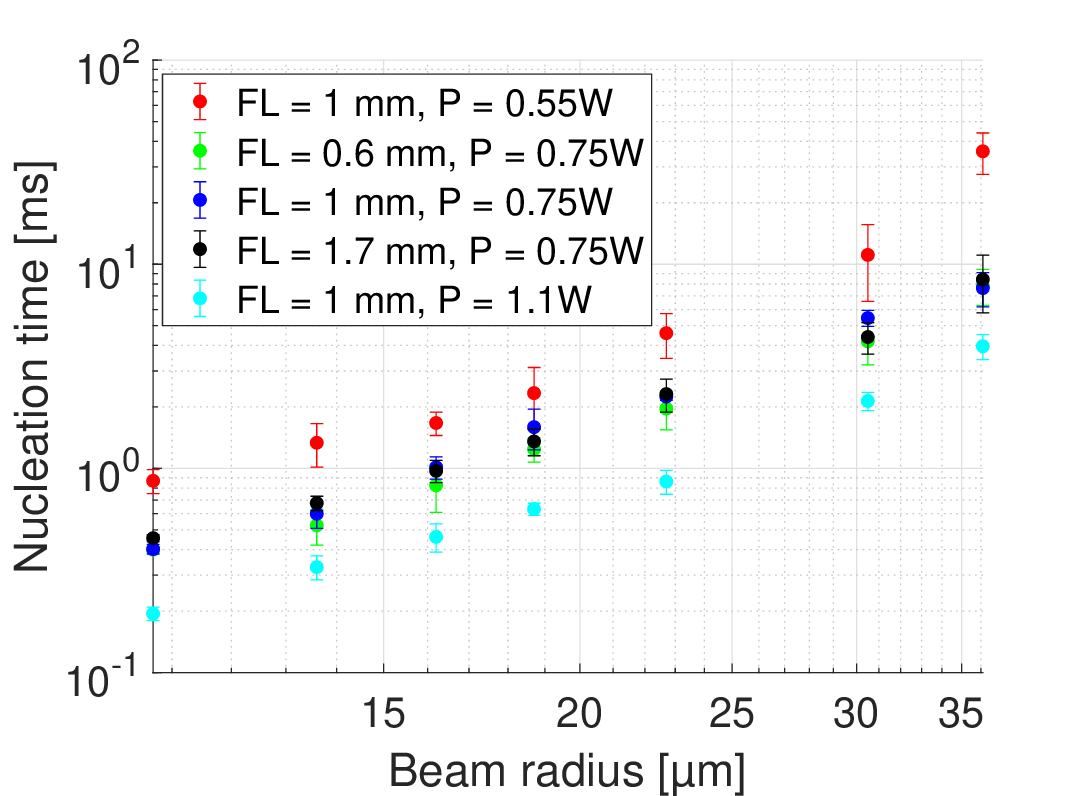}
    \centering
    \captionsetup{width=0.5\linewidth}
    \caption{Experimental nucleation times for three different laser powers and three different filling levels, for a range of beam radii.}
    \label{fig: sup: Nucleation times fill level}
\end{figure} 
\newpage
\section{Simulated nucleation temperatures}\label{sec: suppl. simulated temperatures}
The heat transfer simulations (Section~\ref{sec: suppl. heat transfer}) were executed to find the values and profile of the temperature at the moment of nucleation. Figure~\ref{fig: sup: Nucleation temperatures} shows these maximum simulated temperatures found in the liquid at the experimental nucleation time. The values are approximately constant at 237~$\pm$~10\textdegree C. Only for the smallest laser power (550 mW) and the larger beam radii, the found temperatures are lower. This could be explained by the simulation settings of the glass-liquid interface. At this interface, the temperature on equal sides are equal, which slightly increases the heat dissipation, and therefore slightly decreases the found temperatures, which is especially apparent for the longest nucleation times. 
\begin{figure}[h!]
    \includegraphics[width=0.5\linewidth]{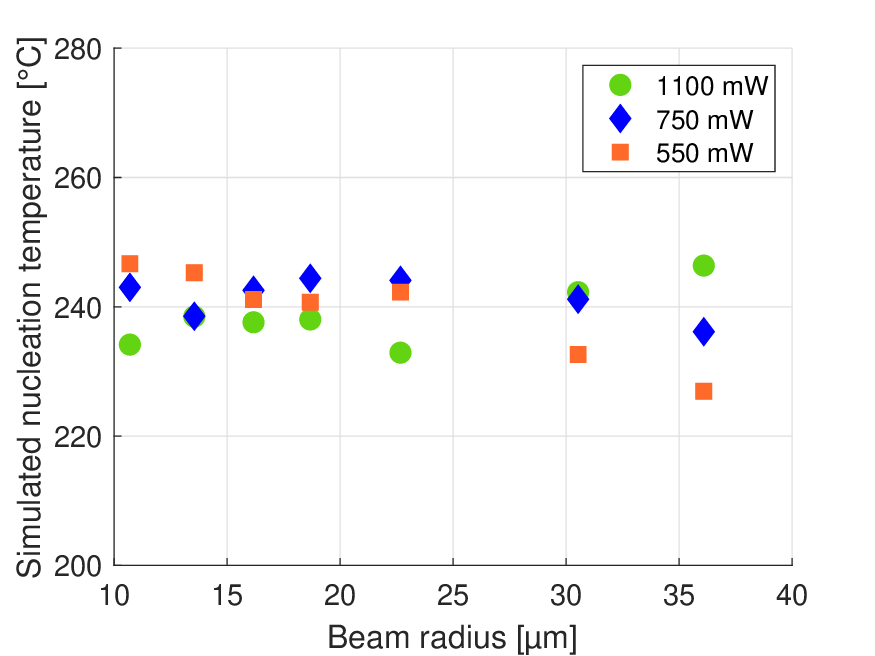}
    \centering
    \captionsetup{width=0.5\linewidth}
    \caption{Maximum temperatures in the simulations at the time equal to the experimentally found nucleation times.}
    \label{fig: sup: Nucleation temperatures}
\end{figure} 
\newpage
\section{Maximum bubble volume}\label{sec: suppl. bubble volume frames}
In most cases, reaches it's maximum volume and then collapses again by condensation of the vapor (see example in Figure~\ref{fig: ExperimentalSetup}). However, for large bubbles, the bubble may grow very large such that it comes in contact with the surrounding air. In these cases, the bubble 'collapses' instantaneously. Such an example is shown in Figure~\ref{fig: sup: Coalesce Collapse}.
\begin{figure}[h!]
    \includegraphics[width=0.5\linewidth]{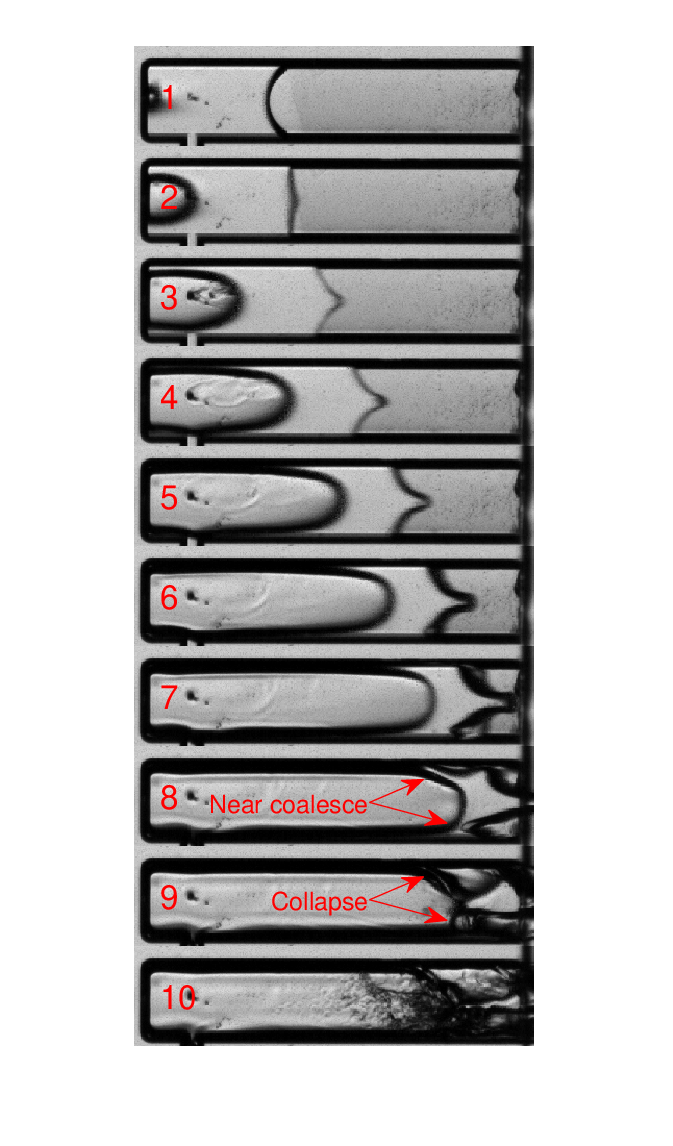}
    \centering
    \captionsetup{width=0.5\linewidth}
    \caption{Series of frames from experiment ($F$~=~600~µm, P~=~0.75~W), showing an example where the bubble coalesces with the surrounding air before reaching its maximum volume, around frame 9. Interframe time is 8.9~µs.}
    \label{fig: sup: Coalesce Collapse}
\end{figure} 
%\addtocontents{toc}{\protect\setcounter{tocdepth}{1}}
\end{supplement}

\end{document}